\documentclass[%
reprint,
superscriptaddress,
amsmath,amssymb,
aps,
]{revtex4-2}

\usepackage[]{graphicx}
\usepackage{commath,amsfonts}
\usepackage{xcolor, ulem}
\usepackage{dcolumn}
\usepackage{bm}
\usepackage{comment}
\newcommand{\iu}{\mathrm{i}\mkern1mu}

\begin{document}

\preprint{APS/123-QED}

\title{Order, chaos, and dimensionality transition in a system of swarmalators}
\author{Joao U. F. Lizárraga}
\affiliation{%
	Instituto de Física Gleb Wataghin, Universidade Estadual de Campinas, Unicamp 13083-970, Campinas, São Paulo, Brazil
}
\author{Kevin P. O'Keeffe}
\affiliation{%
	Senseable City Lab, Massachusetts Institute of Technology, Cambridge, Massachusetts 02139, USA
}
\author{Marcus A. M. de Aguiar}%
\email{aguiar@ifi.unicamp.br}
\affiliation{%
	Instituto de Física Gleb Wataghin, Universidade Estadual de Campinas, Unicamp 13083-970, Campinas, São Paulo, Brazil
}



\date{\today}

\begin{abstract}
	Similar to sperm, where individuals self-organize in space while also striving for coherence in their tail swinging, several natural and engineered systems exhibit the emergence of swarming and synchronization. The arising and interplay of these phenomena have been captured by collectives of hypothetical particles named swarmalators, each possessing a position and a phase whose dynamics are affected reciprocally and also by the space-phase states of their neighbors. In this work, we introduce a solvable model of swarmalators able to move in two-dimensional spaces. We show that several static and active collective states can emerge and derive necessary conditions for each to show up as the model parameters are varied. These conditions elucidate, in some cases, the displaying of multistability among states. Notably, in the active regime, individuals behave chaotically, maintaining spatial correlation under certain conditions, and breaking it under others on what we interpret as a dimensionality transition.
\end{abstract}

\maketitle

\section{Introduction}\label{sec:sec01}
Swarming and synchronization are emergent phenomena observed in various natural systems. Swarming, more noticeable due to its occurrence in physical space, is evidenced in groups of fishes, birds, insects, bacteria, among others~\cite{larsson2012fish, marras2015fish, emlen1952flocking, bialek2012statistical, sullivan1981insect, chittka2022social, kearns2010field, ariel2015swarming, be2019statistical}. Synchronization, sometimes less apparent due to its nature, is observed in both living and non-living systems, including groups of humans, frogs, heart cells, neurons, among others~\cite{neda2000sound, strogatz2005crowd, yaniv2014synchronization, bychkov2020synchronized, aihara2006periodic, aihara2007dynamical, aihara2008mathematical, uhlhaas2006neural}. From a theoretical standpoint, both phenomena have been extensively studied using distinct frameworks based on well-known models named after Kuramoto~\cite{strogatz2000kuramoto}, Stuart-Landau~\cite{aoyagi1995network}, Couzin~\cite{couzin2002collective}, and Vicsek~\cite{vicsek1995novel}. In these models, single particles are represented by their positions or by periodic internal degrees of freedom dubbed phases. Then, individuals' spatial self-organization gives rise to swarming and, in case of phase coherence, to synchronization.

Despite the success in studying swarming and synchronization independently, a significant step up has taken place recently, influencing both fields. A singular type of particles, named swarmalators, have been introduced in~\cite{o2017oscillators} such that, collectively, these can synchronize and swarm as their spatial and phase dynamics interplay. Several systems where individuals' phases influence their positions and vice versa have been spotted in nature. These include magnetic quincke rollers~\cite{zhang2020reconfigurable}, sperm~\cite{creppy2016symmetry, maggi2023thermodynamic, creppy2015turbulence}, starfish embryos~\cite{tan2022odd}, tree frogs~\cite{aihara2014spatio}, nematodes~\cite{peshkov2022synchronized, yuan2014gait}, and more~\cite{erglis2007dynamics, sarfati2020spatio, quillen2022fluid, barotta2023bidirectional, manna2021chemical, ali2023oscillating}.

The swarmalators model, introduced in~\cite{o2017oscillators}, is defined by the equations
\begin{align}
	\begin{split}
		\dot{\vec{p}}_i &= \vec{\varepsilon}_i + \frac{1}{N}\sum_{j = 1}^{N}\left[I_{\alpha}(\vec{p}_{ji})F(\theta_{ji}) - I_{\rho}(\vec{p}_{ji})\right],\\
		\dot{\theta}_i &= \omega_i + \frac{1}{N}\sum_{j = 1}^{N}H_{\alpha}(\theta_{ji})G(\vec{p}_{ji}),
	\end{split}
	\label{eqS1:general}
\end{align}
where $\vec{p}_i$ and $\theta_i$ represent the position and the phase of the $i$-th swarmalator, respectively. Each individual is affected by intrinsic spatial and angular velocities $(\vec{\varepsilon}_i,\omega_i)$, and by the coupling with other individuals, determined by attraction $(I_\alpha, H_{\alpha})$, repulsion $(I_{\rho})$, and influence $(F, G)$ functions. We use $\vec{p}_{ji}$ and $\theta_{ji}$ as compact representations of ($\vec{p}_j - \vec{p}_i$) and ($\vec{\theta}_j - \vec{\theta}_i$), respectively. It was shown, in the same work, that computing a two-dimensional (2D) instance of the model would lead to the emergence of several static and active states, each with unique features. Afterwards, various modifications of this 2D model were introduced, analyzing effects produced by an external stimulus~\cite{lizarraga2020synchronization}, chirality of the particles~\cite{ceron2023diverse}, different interactions~\cite{hong2018active, mclennan2020emergent, sar2022swarmalators}, among others~\cite{belovs2017synchronized,  japon2022intercellular, adorjani2023motility, blum2022swarmalators, kongni2023phase, ghosh2023antiphase}. One drawback shared by most of these studies, however, is that despite their numeric findings, analytic results are very limited given the complexity of the model.

The starting point of the work we present here is the one-dimensional (1D) swarmalators model, also denominated ``ring model'', introduced in~\cite{o2022collective}. Following the same terminology as in Eqs.~\eqref{eqS1:general}, the 1D model is defined by 
\begin{align}
	\begin{split}
		\dot{x}_i &= \frac{1}{N}\sum_{j = 1}^{N}I_\alpha(x_{ji})F(\theta_{ji}),\\
		\dot{\theta}_i &= \frac{1}{N}\sum_{j = 1}^{N}H_\alpha(\theta_{ji})G(x_{ji}), \end{split}
	\label{eqS1:1d_general}
\end{align}
where attraction and influence functions are chosen to be sines and cosines, respectively, weighted by scalar coupling constants, and $(x_i, \theta_i) \in (\mathbb{S}^1, \mathbb{S}^1)$. Individuals are assumed to be identical, so intrinsic velocities are not considered. Thus, the dynamics of the system are defined by a couple of Kuramoto-like equations. In these, the independent synchronization of positions and phases, promoted by the sines, is strengthened by the interplay induced by the cosines. Given the simplicity in handling periodic functions, this version of the 1D swarmalators model is analytically tractable. Taking advantage of this, several solvable variations have been presented afterwards, with a focus on analyzing the effects generated by nonidentical swarmalators~\cite{yoon2022sync}, distributed couplings~\cite{o2022swarmalators}, and other factors~\cite{hao2023mixed, lizarraga2023synchronization, sar2023pinning}. 
One drawback of the 1D swarmalators model in its `bare' form [Eq.~\eqref{eqS1:1d_general}] is that its structure prevents the emergence of active states, where $(\dot{x}_i, \dot{\theta}_i \ne 0)$, in contrast with the 2D instance of Eq.~\eqref{eqS1:general}, where active states emerge even for identical particles. To observe such states in the 1D framework, the solutions found so far involved adding frustration~\cite{lizarraga2023synchronization}, external stimuli~\cite{sar2023pinning}, considering nonidentical individuals~\cite{yoon2022sync}, or by splitting the population and mixing coupling signs~\cite{hao2023mixed}.

Here, we introduce a 2D swarmalators model that shares features with the ones described in Eqs.~\eqref{eqS1:general} and~\eqref{eqS1:1d_general},  leveraging their advantages and overcoming their drawbacks: the particles are able to move in a $2\pi$-periodic 2D space, and the model is tractable analytically.  We show that the model presents several static and active states, some of them similar to ones found in previous studies and some of them new. The active states, moreover, are part of a chaotic regime which, under certain conditions, generates a dimensionality transition.

We outline the general form of our model in Sec.~\ref{sec:sec02}, along with the specific equations used in this work. Next, in Sec.~\ref{sec:sec03}, we present the static and active states that emerge through numerical computations of the model. Upon the intuition gained from the numerical results, in Sec.~\ref{sec:sec04}, we show the main results concerning the stability of the static states. In Sec.~\ref{sec:sec05}, we present the conditions where the system shows chaotic behaviors and the dimensionality transition. In Sec.~\ref{sec:sec06}, we summarize static and active states constructing diagrams showing their stable regions. We conclude this work by discussing our findings and future research suggestions in Sec.~\ref{sec:sec07}.

\section{The model}\label{sec:sec02}
We aim to exploit the solvability of the 1D model introduced in~\cite{o2022collective}, in a 2D setup. However, before we describe our generalization, we find it useful to comment on the limitations that arise when trying to recover the dynamics in Eqs.~\eqref{eqS1:general} from Eqs.~\eqref{eqS1:1d_general}. First, and most evident, the 1D model lacks an explicit repulsive component. This drawback hinders the emergence of states that rely on the different scales of attraction and repulsion forces. For instance, the 1D model cannot generate active states, where the particles keep moving after a transient. Even if an additional position coordinate $y_i$ is considered, defined to be symmetric with $x_i$, the Kuramoto-like feature, as defined in~\cite{o2022collective}, would lead to the synchronization in each axis, collapsing the system to static analogs of the 1D states. Moreover, the attraction and repulsion functions in Eqs.~\eqref{eqS1:general} depend on the spatial distances between the particles $\vec{p}_{ji}$. For example, the 2D instance presented in~\cite{o2017oscillators} use a power law in $p_{ji}$ as repulsion function, generating collective states where particles distribute radially. Extensions of the ring model to 2D must, therefore, include explicit dependence of repulsive interactions on particle's positions. 

For our 2D setup, we define the position of the $i$-th swarmalator as $\vec{p}_i = (x_i, y_i)$ but use the distances in each axis, $x_{ij}$ and $y_{ij}$, independently in the dynamical equations, as using the modulus $p_{ij}$ would undermine the reduction of complexity that we are looking for. 
The coupling between $x$ and $y$ is done through the phases $\theta_i$, included in the functions $F_\alpha$ and $F_\rho$ that affect attraction $I_\alpha$ and repulsion $I_\rho$, respectively.  The phase dynamics, on the other hand, are defined by their mutual interaction $H$ and by the interplay of both spatial coordinates $G$. In its general form the model is described by
\begin{align*}
	\begin{split}
		\dot{x}_i &= u_i+ \frac{1}{N}\sum_{j = 1}^{N}\left[I_\alpha(x_{ji})F_\alpha(\theta_{ji}) - I_\rho(x_{ji})F_\rho(\theta_{ji})\right],\\
		\dot{y}_i &= \upsilon_i+ \frac{1}{N}\sum_{j = 1}^{N}\left[I_\alpha(y_{ji})F_\alpha(\theta_{ji}) - I_\rho(y_{ji})F_\rho(\theta_{ji})\right],\\ 
		\dot{\theta}_i &= \omega_i + \frac{1}{N}\sum_{j = 1}^{N}H_\alpha(\theta_{ji})G(x_{ji}, y_{ji}). \end{split}
\end{align*}
For the specific model we study here, we consider that $(x_i, y_i, \theta_i)\in(\mathbb{S}^1, \mathbb{S}^1, \mathbb{S}^1)$, and individuals are assumed to be identical, so $(u_i, \upsilon_i, \omega_i) = 0$. Following~\cite{o2022collective}, we choose combinations of sines and cosines for all functions, weighted by coupling constants $J^\pm$ and $K'$. This makes the problem amenable to analytical treatment, but introduces repulsion in a weak sense: 

\begin{widetext}
	\begin{align}
		\begin{split}
			\dot{x}_i &= \frac{1}{N}\sum_{j= 1}^N \left\{J^-\sin (x_{ji})\cos(\theta_{ji}) - J^+\left[1- \cos (x_{ji})\right] \sin(\theta_{ji}) \right\},\\
			\dot{y}_i &= \frac{1}{N}\sum_{j= 1}^N \left\{J^-\sin (y_{ji})\cos(\theta_{ji}) - J^+\left[1- \cos (y_{ji})\right] \sin(\theta_{ji}) \right\},\\
			\dot{\theta}_i &= \frac{K'}{N}\sum_{j= 1}^N \sin(\theta_{ji})\left[2+ \cos(2x_{ji}) + \cos(2y_{ji})\right].
		\end{split}
		\label{eqS2:2d_general}
	\end{align}
\end{widetext}

In this form, $J^-$ is the weight of the Kurmamoto-like attractive interaction, that is enhanced by phase synchronization. $J^+$, on the other hand, is the weight of the weak repulsion term, whose sign depends on the difference of phases. It is weak because it disappears when particles are on top of each other, allowing for full synchronization in $x$, $y$ and $\theta$. When phases are synchronized, $\theta_{ij}=0$, the equations for $x_i$ and $y_i$ reduce to independent Kuramoto dynamics. Similarly, when $x_{ij}=y_{ij}=0$, the equation for $\theta_i$ follows the Kuramoto model. Repulsion and attraction terms in each direction depend only on the distances in that direction. This is the key feature that allows for simplifications in the analytical treatment while still supporting active states.

As a final step, for an easier understanding of the system, we rewrite $J^{\pm}~=~(J_A \pm J_R)/2$ and $K' = K/2$. These new parameters do not alter the structure shown in Eqs.~\eqref{eqS2:2d_general}, but facilitate manipulating attraction and repulsion scales, affected respectively by $J_A$ and $J_R$. Thus, the expanded model is given by
\begin{widetext}
	\begin{align}
		\begin{split}
			\dot{x}_i &= \frac{1}{N}\sum_{j = 1}^N\frac{J_A}{2}\left\{\sin(x_{ji})\cos(\theta_{ji}) - \left[1- \cos(x_{ji})\right]\sin(\theta_{ji})\right\} - \frac{J_R}{2}\left\{\sin(x_{ji})\cos(\theta_{ji}) + \left[1- \cos(x_{ji})\right]\sin(\theta_{ji}) \right\},\\
			\dot{y}_i &=  \frac{1}{N}\sum_{j = 1}^N\frac{J_A}{2}\left\{\sin(y_{ji})\cos(\theta_{ji}) - \left[1- \cos(y_{ji})\right]\sin(\theta_{ji})\right\} - \frac{J_R}{2}\left\{\sin(y_{ji})\cos(\theta_{ji}) + \left[1- \cos(y_{ji})\right]\sin(\theta_{ji}) \right\},\\ 
			\dot{\theta}_i &= \frac{K}{N}\sum_{j = 1}^N \sin(\theta_{ji})\left\{1 + \frac{1}{2}\left[\cos(2x_{ji}) + \cos(2y_{ji})\right]\right\}.
		\end{split}
		\label{eqS2:expanded}
	\end{align}
\end{widetext}

Notice that, in this form, the terms proportional to $J_R$ in the equation for $\dot{x}_i$ include a truly  repulsive interaction, $-\sin(x_{ji})$, and half of the weakly repulsive term,  $-[1-\cos(x_{ji})]$ (and similarly for $\dot{y}_i$). The terms 
in $J_A$, however, include the attractive part $\sin(x_{ji})$ but also the other half of the weakly repulsive term. By considering this expanded 2D model, we can derive the original ring model under intuitive convenient conditions (see Appendix~\ref{app:app01}).

\section{Numerical simulations}\label{sec:sec03}
\begin{figure*}
	\centering
	\includegraphics[width = 0.8\textwidth]{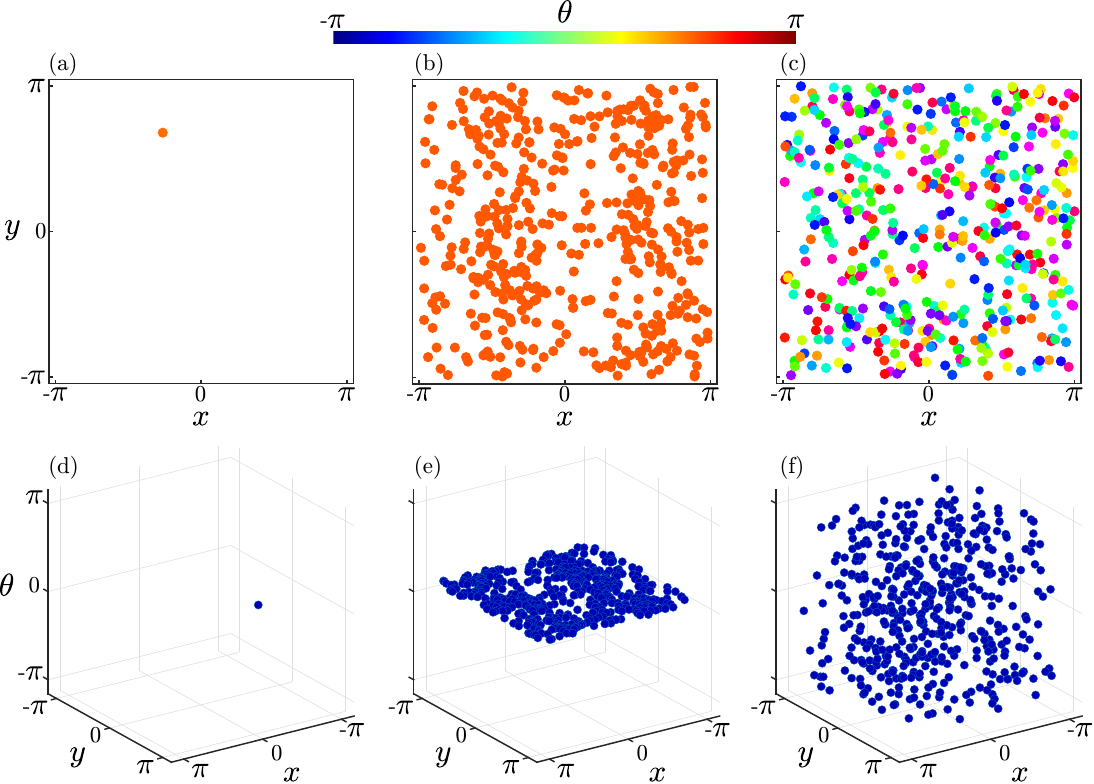}
	\caption{Snapshots of the [(a),(d)] point synchronous, [(b),(e)] distributed synchronous, and [(c),(f)] distributed asynchronous states after a transient. The panels show the spatial distribution of the particles (top row) and the respective correlation scatter plots (bottom row). The parameters $(K, J_A, J_R)$ are set as $(1, 1, 0.5)$ for the state in [(a), (d)], $(1, 0.5, 1)$ for the state in [(b), (e)], and $(0, -0.5, 0.5)$ for the state in [(c),(f)]. See movies S1, S2, and S3 in the Supplemental Material.}
	\label{fig:states01}
\end{figure*}

\begin{figure*}
	\centering
	\includegraphics[width = 0.8\textwidth]{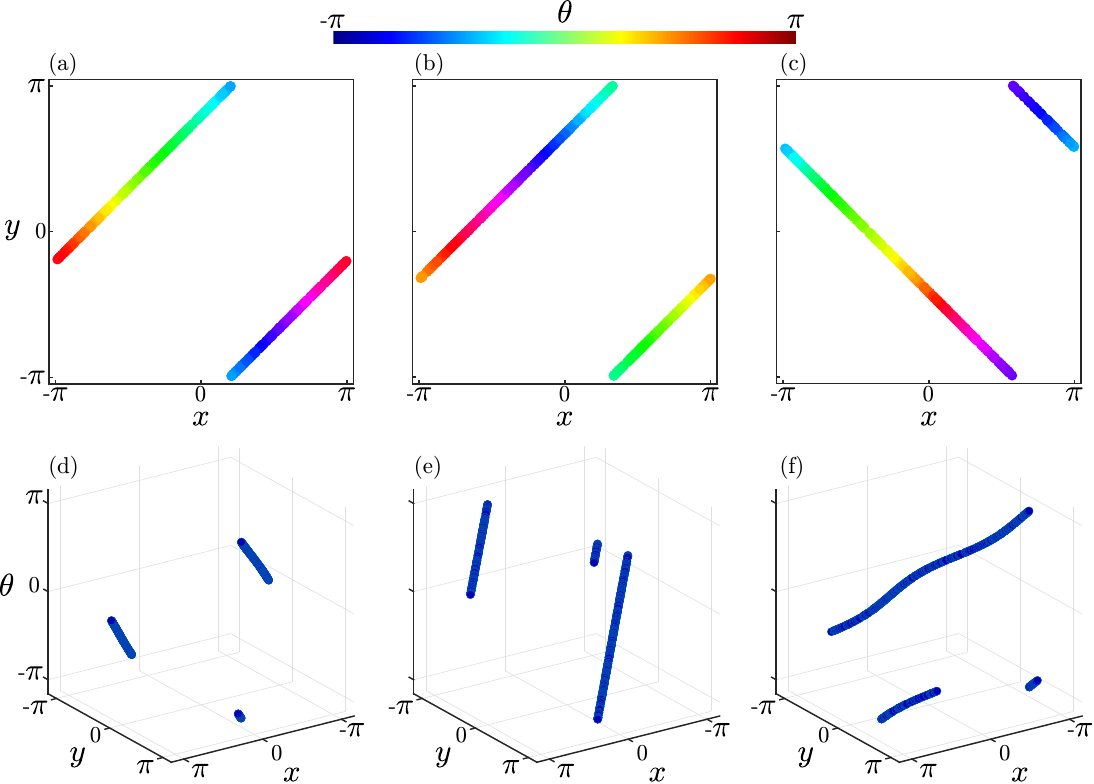}
	\caption{Snapshots of the static phase wave states after a transient. The panels show the spatial distribution of the particles (top row) and the respective correlation scatter plots (bottom row).[(a), (d)], [(b), (e)], and [(e), (f)] pairs correspond to different types of correlation between $x_i, y_i$, and $\theta_i$ states. The parameters $(K, J_A, J_R)$ are set as $(0, 0, -0.8)$ for the state in [(a), (d)], $(0, 0.8, 0)$ for the state in [(b), (e)], and $(0, 1, -1)$ for the state in [(c),(f)]. See movies S4, S5, and S6 in the Supplemental Material.}
	\label{fig:states02}
\end{figure*}

\begin{figure*}
	\centering
	\includegraphics[width = 0.8\textwidth]{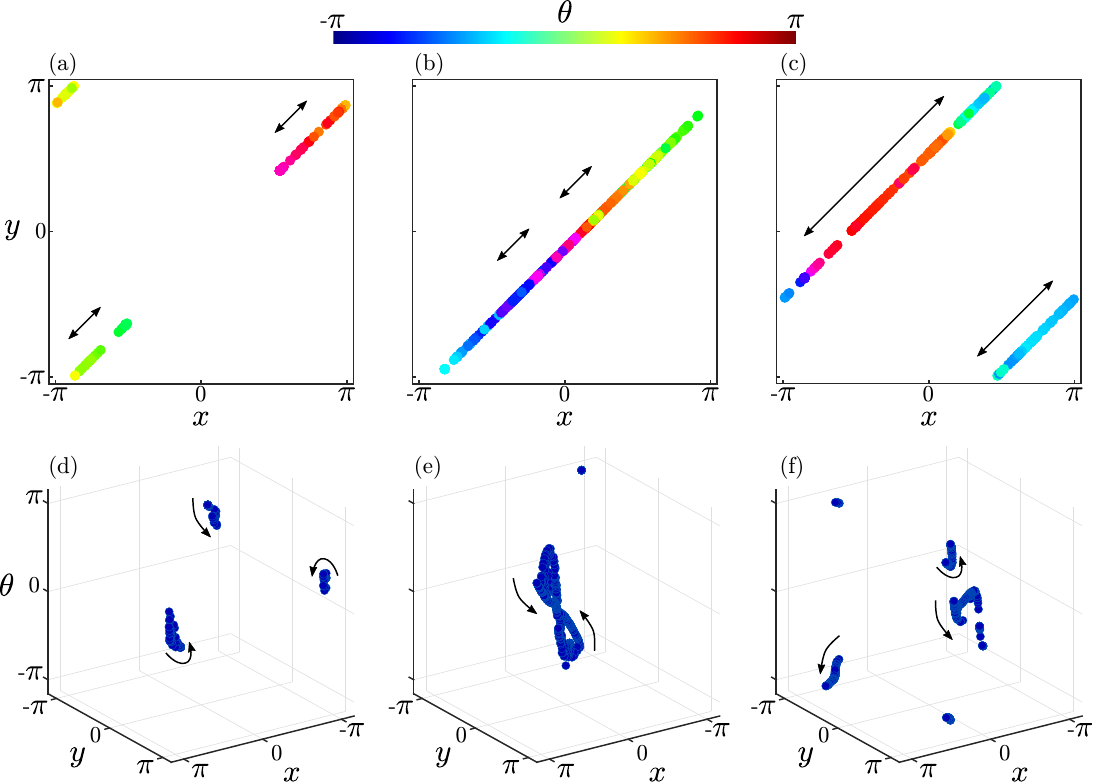}
	\caption{Snapshots of the [(a),(d)] swirling, [(b),(e)] butterfly, and [(c),(f)] bouncing states after $10^5$ time-steps. The panels show the spatial distribution of the particles (top row) and the respective correlation scatter plots (bottom row). The arrows represent the motion of the particles. The parameters $(K, J_A, J_R)$ are set as $(-1, 0, -0.5)$ for the state in [(a), (d)], $(-1, -0.5, -0.5)$ for the state in [(b), (e)], and $(-1, -1, -0.5)$ for the state in [(c),(f)]. See movies S7, S8, and S9 in the Supplemental Material.}
	\label{fig:states03}
\end{figure*}

Simulations performed for the model described in Eqs.~\eqref{eqS2:expanded} show the emergence of states that gather features of several states observed in instances of Eqs.~\eqref{eqS1:general} and~\eqref{eqS1:1d_general}. Additionally, we observe active states that, despite the similarity with states presented in previous works, show interesting properties. All these states are presented in Figs.~\ref{fig:states01},~\ref{fig:states02}, and~\ref{fig:states03}, where the computations were performed for a population of $N = 500$ swarmalators and different values of $J_A$, $J_R$, and $K$. The particles start positioned uniformly in a $2\pi$-length cube in the $(x,y,\theta)$ space, and their states evolve along $10^5$ time-steps. From Fig.~\ref{fig:states01} to Fig.~\ref{fig:states03}, each shows the particles distributed in the $(x, y)$ space colored according to their phases (top rows), and their respective distributions in the $(x, y, \theta)$ space (bottom rows).

In Figs.~\ref{fig:states01}(a) and~\ref{fig:states01}(d), we present the point synchronous state, which shows the phase synchronization of particles and their convergence to a fixed point in the $(x, y)$ space. Once these reach a $(x_0, y_0, \theta_0)$ state, they remain steady. A similar state emerges on the 1D model~\cite{o2022collective}, where particles converge to a fixed position on the ring. In our model, the emergence of this state requires phase synchronization $(K > 0)$, and also $(J_A, J_R > 0)$. Moreover, attraction must be stronger than repulsion, $J_A > J_R$, so that particles, which start distributed across the $(x, y)$ space, can collapse to a fixed position.  For $J_R > J_A$ the particles converge to the distributed synchronous state shown in Figs.~\ref{fig:states01}(b) and~\ref{fig:states01}(e). In this state, particles synchronize (as $K>0$) but they remain distributed in the $(x,y)$ space given that now repulsion is stronger than attraction. This state shows a first approach to spatial states generated by instances of Eqs.~\eqref{eqS1:general}, where particles tend to distribute radially~\cite{o2017oscillators, hong2018active, lizarraga2020synchronization}. Finally, in Figs.~\ref{fig:states01}(c) and~\ref{fig:states01}(f), we show the distributed asynchronous state. In this, the individuals keep their initial phases frozen $(K = 0)$, and the repulsion between them is strong enough to keep the particles distributed uniformly in the $(x,y)$ space. Similar states can also be observed emerging in the 1D model~\cite{o2022collective, o2022swarmalators}, where particles remain asynchronized while distributed along the ring, and in 2D and 3D instances of Eqs.~\eqref{eqS1:general}, where the spatial distribution is radial~\cite{o2017oscillators, hong2018active}.

Panels in Fig.~\ref{fig:states02} show three types of static phase wave state. First, Figs.~\ref{fig:states02}(a) and~\ref{fig:states02}(d) show that individuals' positions $(x_i, y_i)$ are positively correlated and that phases $\theta_i$ are also positively correlated to these positions. Second, Figs.~\ref{fig:states02}(b) and~\ref{fig:states02}(e) show that individuals' positions $(x_i, y_i)$ are positively correlated but that phases $\theta_i$ are negatively correlated to $x_i$ and therefore $y_i$. Third, Figs.~\ref{fig:states02}(c) and~\ref{fig:states02}(f) show that individuals' positions $(x_i, y_i)$ are negatively correlated and that phases $\theta_i$ are positively correlated to $x_i$ only. Each of these states shows similarities with states found in previous work. The static phase wave state in 1D shows the emergence of correlations between particles' positions and phases~\cite{o2022collective}, and in 2D, the correlation emerges between the polar angle that describes the radial position of each particle and its respective phase~\cite{o2017oscillators}. A remarkable feature in our model, furthermore, is that independent of the type of correlation, it features the concept of \textit{`like attracts like'} introduced in~\cite{o2017oscillators}: despite individuals' phases being frozen $(K = 0)$, they end up grouping in space with similar phases individuals. For simplicity, in the following sections, we will refer to each type of static phase wave state as summarized in Table~\ref{tab:tab01}. An additional observation is that, Figs.~\ref{fig:states02}(c) and~\ref{fig:states02}(f) show a small curvature in the correlation, that reminds us of the buckled phase wave state introduced in~\cite{o2022collective}.

\begin{table}
	\begin{tabular}{ccc}
		\hline
		& \multicolumn{2}{c}{Type of correlation}                \\ \hline
		State                 & Positive                    & Negative                 \\ \hline
		Static phase wave I   & $x_i, y_i,\theta_i$ &                          \\ \hline
		Static phase wave II  & $x_i, y_i$                        & $x_i,\theta_i$ \\ \hline
		Static phase wave III & $x_i,\theta_i$    & $x_i, y_i$                     \\ \hline
	\end{tabular}
	\caption{Classification of the static phase wave states according to the type of correlation between individuals' positions and phases.}
	\label{tab:tab01}
\end{table}

In Fig.~\ref{fig:states03}, we show instances of the most noteworthy states of this work: the active ones. We call these `instances', since all of them are part of a chaotic regime that we explore in more detail in the following sections. The first one, presented in Figs.~\ref{fig:states03}(a) and~\ref{fig:states03}(d), shows that after a transient, there is an apparent correlation between $x_i$, $y_i$, and $\theta_i$, similar to the ones presented for the static phase wave states in Fig~\ref{fig:states02}. However, individuals move, and despite the positive correlation between $x_i$ and $y_i$, shown in Fig.~\ref{fig:states03}(a), individuals' positions and phases generate unsteady irregular annular shapes, as noted in Fig.~\ref{fig:states03}(d). These swirls remind us of the ones formed by non-conformist individuals in~\cite{hao2023mixed}, so we name our state after it. We refer to the second instance of active state, shown in Figures~\ref{fig:states03}(b) and~\ref{fig:states03}(e), as butterfly state. In this state, particles move while positions $x_i$ and $y_i$ keep linearly correlated. Furthermore, the population of particles continuously alternates between splitting into two clusters and merging back to a single one. Similar to the swirling state, individuals' positions and phases are not correlated. Furthermore, the shape generated by these, as shown in Fig.~\ref{fig:states03}(e), reminds us of the well-known Lorenz attractor projected in 2D. Finally, the bouncing state, depicted in Figs.~\ref{fig:states03}(c) and~\ref{fig:states03}(f), shows similarities with the butterfly state in terms of the correlation between individuals' positions [Fig.~\ref{fig:states03}(c)] and the shapes generated between their positions and phases [Fig.~\ref{fig:states03}(f)]. However, the population of particles goes back and forth, from a single cluster to several ones, in an erratic way. Despite this behavior, the linear correlation between $x_i$ and $y_i$ is preserved.

\section{Static states}\label{sec:sec04}
We perform perturbation analyses to study the stability of the static equilibrium states reached by our system. The computations we carry out are based on the analyses presented in~\cite{lizarraga2023synchronization} for an instance of the 1D model. Then, since the procedures are known, only major considerations and results are presented in the main text and complex calculations are left for the Appendix.

\subsection{Point synchronous (PS) state }\label{sub:PS}
This state is characterized by the convergence of the particles to a point state $x_i = x$, $y_i = y$, and $\theta_i = \theta$, which proves to be a solution of Eqs.~\eqref{eqS2:2d_general}. We add small perturbations $\delta x_i$, $\delta y_i$, and $\delta \theta_i$ to each particle in their equilibrium state, and find expressions for their time evolution. The perturbation dynamics and their analysis as a linear system are shown in detail in Appendix~\ref{app:app02}. 

The eigenvalues that determine the stability of this state are
\begin{align}
	\begin{split}
		\lambda^{PS}_0 &= 0,\\
		\lambda^{PS}_1 &= -2K,\\
		\lambda^{PS}_2 &= \frac{J_R - J_A}{2},
	\end{split}
	\label{eqS4:eigsPSS}
\end{align}
revealing that the PS state emerges only when $K > 0$ and $J_A\geq J_R$, as expected from the remarks in the previous section: phase synchronization is driven by $K>0$ and particles' clustering result from $J_A > J_R$. Moreover, from Eqs.~\eqref{eqS2:2d_general}, we can see that phase synchrony strengthen the attractive effects in spatial dynamics, reason why particles converge to a single point in space instead of cluttering while keeping distance from each other. 

\subsection{Static phase wave (SPW) states}\label{sub:SPW}
The main feature shared by these states is the linear correlation among $x_i$, $y_i$, and $\theta_i$, as classified in Tab.~\ref{tab:tab01}. It is easy to check that equilibrium states based on these correlations satisfy the dynamical equations. 

In the static phase wave I, the equilibrium state is defined by $x_i = 2\pi i/N + x_0$, $y_i = 2\pi i/N + y_0$, and $\theta_i = 2\pi i/N + \theta_0$. The perturbation analysis performed for this equilibrium is described in Appendix~\ref{app:SPW1}. The eigenvalues that determine the stability of this state are

\begin{align}
	\begin{split}
		\lambda^{(I)}_0 &= 0, \\
		\lambda^{(I)}_1 &= 0.5J_R, \\    
		\lambda^{(I)}_2 &= 0.5J_R + 0.25J_A, \\
		\lambda^{(I)}_3 &= 0.25J_R + 0.125K \\
		& \qquad \pm 0.5(0.25J_R^2 + 0.0625K^2 - 1.25J_RK )^{0.5},\\
		\lambda^{(I)}_4 &= 0.25J_R + 0.375K \\
		& \qquad \pm 0.5(0.25J_R^2 + 0.562K^2 + 0.75J_RK \\
		&\qquad \qquad \qquad\qquad\qquad\qquad\qquad + 0.5J_AK )^{0.5}.\\
	\end{split}
	\label{eqS4:eigsSPW1}
\end{align}

The simplest case to frame, using these eigenvalues, is when $(K = 0)$, which implies that individual phases will remain distributed randomly as in their initial state. Then, the conditions for stability are $J_R\leq 0$ and $J_R\leq-0.5J_A$. The negative repulsion and the positive attraction terms indicate a primarily attractive nature of the particles. However, because phases are different, particles cluster with others of similar phases. Additionally, considering $(K>0)$ would lead to phase synchronization and, consequently, disrupt the clustering behavior among different phases. This behavior is also aligned with our expectation for this state based on the nature of our model. The case where $(K<0)$ is more complex given that the phase asynchrony is now weighted, incorporating an active ingredient to the system. We explore this active state in the following section.

The static phase wave II is given by $x_i = 2\pi i/N + x_0$, $y_i = 2\pi i/N + y_0$, and $\theta_i = -2\pi i/N + \theta_0$. The details of the perturbation analysis for this equilibrium state are shown in Appendix~\ref{app:SPW2}. The eigenvalues that determine its stability are

\begin{align}
	\begin{split}
		\lambda^{(II)}_0 &= 0, \\
		\lambda^{(II)}_1 &= -0.5J_A, \\    
		\lambda^{(II)}_2 &= -0.5J_A - 0.25J_R, \\
		\lambda^{(II)}_3 &=  -0.25J_A + 0.125K\\
		&\qquad \pm 0.5(0.25J_A^2 + 0.0625K^2 + 1.25J_AK )^{0.5},\\
		\lambda^{(II)}_4 &=  -0.25J_A + 0.375K \\
		&\qquad \pm 0.5(0.25J_A^2 + 0.562K^2 - 0.75J_AK\\
		&\qquad \qquad \qquad\qquad\qquad\qquad\qquad  - 0.5J_RK )^{0.5}.\\
	\end{split}
	\label{eqS4:eigsSPW2}
\end{align}

The conditions for stability are similar to those found in the previous case (static phase wave I): for $K = 0$, we need $J_A \geq 0$ and $-J_A \geq 0.5J_R$.  Both conditions suggest the attractive nature of the particles, which will lead to the clustering of particles with similar phases. When $K > 0$, diversity among phases is disrupted, whereas $K < 0$ drives the emergence of active states. The main difference between this state and the previous one lies in the regions of the $(J_A-J_R)$ plane where each state appears, as we will demonstrate in a following section. 

Finally, the equilibrium states in the static phase wave III is defined as $x_i = 2\pi i/N + x_0$, $y_i = -2\pi i/N + y_0$, and $\theta_i = 2\pi i/N + \theta_0$. Despite the similarity of these expressions to the equilibria in the two previous cases, the analysis is more complex. The negative correlation between $x_i$ and $y_i$ prevent simplifications in the perturbation analysis. Consequently, finding the eigenvalues that characterize the state's stability becomes more intricate. For details of this analysis, we refer the reader to the Appendix~\ref{app:SPW3}. Although we do not present the corresponding eigenvalues in the main text, in a following section we will show the stability regions of this state in the $(J_A-J_R)$ plane.

\subsection{Distributed states}\label{sub:DS}
To study the stability of these states we assume a continuum of particles, instead of a discrete set like in the previous cases. We define the fraction of particles lying between $x$, $y$, $\theta$ and $x+\mathrm{d}x$, $y+\mathrm{d}y$, $\theta + \mathrm{d}\theta$ at time $t$, by the density $\rho(x, y, \theta, t)\mathrm{d}x \mathrm{d}y \mathrm{d}\theta$. Moreover, $\rho$ will satisfy the normalization condition
\begin{equation}
	\int \rho(x, y, \theta) \mathrm{d}x \mathrm{d}y \mathrm{d}\theta = 1
	\label{eqS4:norm}
\end{equation}
where all integrals run from 0 to $2\pi$.

To perform the stability analysis we introduce order parameters of the form
\begin{equation}
	S_\sigma e^{\iu\phi_{\sigma}} = \frac{1}{N}\sum_{j = 1}^N e^{\iu\sigma_j}.
	\label{eqS4:OPs}
\end{equation}
In this expression, $\sigma$ can represent any individual variable $x$, $y$, $\theta$, or a linear combination of these. So, for instance, if we consider $\sigma:\theta$, the order parameter will measure phase coherence, giving $S_{\theta} = 1$ for a fully synchronized state. Another instance, a bit more interesting is the consideration of $\sigma:(\theta + x)$, which measures the correlation between individuals' positions $x_j$ and phases $\theta_j$. In this case, getting $S_{(\theta + x)} = 1$ is an indicator of a negative correlation between $x_i$ and $\theta_i$ as in the static phase wave state II. 

In the continuum limit, considering Eq.~\eqref{eqS4:norm}, we can rewrite the order parameters as

\begin{equation}
	S_{\sigma}e^{\iu\phi_{\sigma}}=\int_{0}^{2\pi}e^{\iu\sigma} \rho(x, y, \theta) \mathrm{d}x \mathrm{d}y \mathrm{d}\theta,
\end{equation}
so that Eq.~\eqref{eqS2:2d_general} turns into the mean-field equations
\begin{widetext}
	\begin{align}
		\begin{split}
			v_x &= \frac{J_A}{2}S_{(\theta + x)}\sin\left(\phi_{(\theta + x)} - \theta-x\right) + \frac{J_R}{2}S_{(\theta - x)}\sin\left(\phi_{(\theta - x)} - \theta + x\right) - \frac{J_A + J_R}{2}S_\theta\sin\left(\phi_{\theta} - \theta\right),\\   
			v_y &= \frac{J_A}{2}S_{(\theta + y)}\sin\left(\phi_{(\theta + y)} - \theta - y\right) + \frac{J_R}{2}S_{(\theta - y)}\sin\left(\phi_{(\theta - y)} - \theta + y\right) - \frac{J_A + J_R}{2}S_\theta\sin\left(\phi_{\theta} - \theta\right),\\
			v_{\theta} &= KS_{\theta}\sin\left(\phi_{\theta} - \theta\right)+\frac{K}{4}\Big[S_{(\theta + 2x)}\sin\left(\phi_{(\theta + 2x)} - \theta - 2x\right) + S_{(\theta - 2x)}\sin\left(\phi_{(\theta - 2x)} - \theta + 2x\right)\\
			&\qquad \qquad \qquad \qquad \qquad \qquad + S_{(\theta + 2y)}\sin\left(\phi_{(\theta + 2y)} - \theta - 2y\right)+ S_{(\theta - 2y)}\sin\left(\phi_{(\theta - 2y)} - \theta + 2y\right)\Big].
		\end{split}
		\label{eqS4:dynField}
	\end{align}
\end{widetext}

A fully incoherent state, as depicted in the distributed asynchronous case, would be portrayed by the convergence of all the order parameters, in Eqs.~\eqref{eqS4:dynField}, to zero. However, we can also describe the distributed synchronous state by considering that incoherence happens in space only, letting the system reach phase synchronization ($S_{\theta} = 1$). For both cases, we define the respective incoherent densities  
\begin{align}
	\begin{split}
		\rho_0^A &= \frac{1}{8\pi^3}, \\
		\rho_0^S &= \frac{1}{4\pi^2}, 
		\label{eqS4:incohEq}
	\end{split}
\end{align}
which under conditions of the order parameters described above, prove to be equilibrium states of the continuity equation 
\begin{equation}
	\frac{\partial\rho}{\partial t}  = - \vec{\nabla}(\rho \vec{v}), 
	\label{eqS4:contiEq}
\end{equation}
where $\vec{v} = (v_x, v_y, v_{\theta})$ is the velocity field defined in Eq.~\eqref{eqS4:dynField}. We perform perturbation analyses for both incoherent states (see Appendix~\ref{app:app03}). The eigenvalues that determine the stability of the distributed asynchronous state are
\begin{align}
	\begin{split}
		\lambda_1^{A} & = \frac{J_A}{16\pi^2},\\
		\lambda_2^{A} & = -\frac{J_R}{16\pi^2},\\
		\lambda_3^{A} & = \frac{K}{8\pi^2},\\
		\lambda_4^{A} & = \frac{K}{32\pi^2},
	\end{split}
	\label{eqS4:eigs_SA}
\end{align}
and, for the distributed synchronous state, 
\begin{equation}
	\lambda^{S}  = \frac{J_A- J_R}{8\pi^2},
	\label{eqS4:eigs_SS}
\end{equation}
which additionally requires $(K>0)$ to reach phase coherence.
\section{Active states}\label{sec:sec05} 
In the previous section, we discussed all the conditions that allow the emergence of the static states. Despite these analytical results, we found only very few hints on the conditions driving the emergence of active states. In order to get more clues on their behavior, we start our analysis by considering distributed states with finite number of particles. Then,  we will demonstrate that all the active states shown in Sec.~\ref{sec:sec03} are actually part of the same chaotic regime. 

\subsection{Active asynchronous state}

The analysis of distributed states performed in Sec.~\ref{sub:DS} relied on the assumption that the number of particles is infinite, so that we could take the continuum limit. This allowed us to consider that all order parameters in Eq.~\eqref{eqS4:dynField} would converge asymptotically to zero. However, when dealing with a finite number of particles, this last assumption is not quite true. For $K=0$, in particular (frozen phases), phase incoherence is not perfect and $S_\theta>0$. This drawback promotes the appearance of small velocities in $\dot{x}_i$ and $\dot{y}_i$, proportional to $(J_A + J_R)/2$, which ultimately will drive the emergence of an active state. For an instance of this state, see movie S10 in the Supplemental Material. If $J_A = - J_R$, i.e., attraction and repulsion have the same intensity, the state still converges to a static configuration.

\subsection{Chaos in 3N dimensions}\label{sub:3NChaos}
In Sec.~\ref{sec:sec04} we studied the stability of static states by computing the eigenvalues of the linearized dynamics in scenarios where phases were frozen, $K = 0$, or driven towards synchronization by $K > 0$. Negative phase couplings ($K<0$), however, not only makes computations more complex but also drives the emergence of active states (as shown in Figs.~\ref{fig:states03}). 

\begin{figure}
	\centering
	\includegraphics[width = \columnwidth]{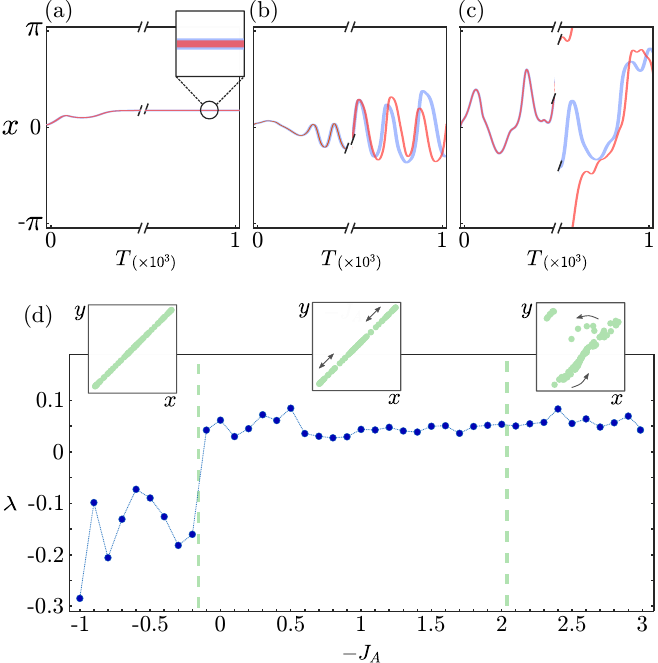}
	\caption{Temporal behavior of single particle trajectories along the $x$-axis with non-perturbed (lavender) and perturbed (red) initial conditions (top row), and maximum Lyapunov exponents for different values of $J_A$ (bottom row). Parameters $(K, J_R)$ are set as $(-1, -0.5)$ for all the figures. Trajectories are generated for (a) $J_A = 0.5$, (b) $J_A = -0.5$, and (c) $J_A = -3$. The square in the top corner of (a) shows a magnification of the circled trajectories, and the ones in (d) show correlations in $x-y$ coordinates depending on the values of $J_A$ (See movies S11 and S12 in the Supplemental Material). The green dashed lines represent the change from static phase wave to chaotic states (left) and the breaking of $x-y$ coordinates correlation (right).}
	\label{fig:states04}
\end{figure}

An interesting feature of these states is that particles are sensitive to small changes in their initial conditions. We highlight this property in Figs.~\ref{fig:states04}(a), ~\ref{fig:states04}(b), and~\ref{fig:states04}(c), where trajectories in the $x$-axis are shown for the same particle in scenarios where $K = -1$. In each of these, we compare the trajectories followed by a single particle when the system starts on the $N$-dimensional initial states $(\vec{x}_0, \vec{y}_0, \vec{\theta}_0)$ and $(\vec{x}_0+\delta^{(1)}_x, \vec{y}_0, \vec{\theta}_0)$, where $\delta^{(1)}_{x} =10^{-4}$ represents that only the particle of interest's state is perturbed (for instance $x_{1(0)} + \delta^{(1)}_x$). It is clear then that, when considering $(J_A, J_R) = (0.5, -0.5)$ [Fig.~\ref{fig:states04}(a)], the perturbation does not affect the trajectory of the particle considerably. In fact, from Eqs.~\eqref{eqS4:eigsSPW1}, we can infer that these parameters drive the emergence of the static phase wave I. However, once $J_A$ becomes negative [$(J_A, J_R) = (-0.5, -0.5)$ for Fig.~\ref{fig:states04}(b) and $(J_A, J_R) = (-0.5, -3)$ for Fig.~\ref{fig:states04}(c)], the perturbed trajectories suggest the existence of chaos.

In order to explore the presence of chaotic states we compute the maximum Lyapunov exponent as a function of $J_A$ for $K=-1$ and $J_R=-0.5$. We take two $3N$-dimensional trajectories of the system, one with initial conditions $(\vec{x}_0, \vec{y}_0, \vec{\theta}_0)$, and the other one with $(\vec{x}_0+\delta^{(1)}_x, \vec{y}_0, \vec{\theta}_0)$. We remark that, although $\delta^{(1)}_x$ affects the initial condition of a single particle in the $x$ direction, it perturbs the entire $3N$-dimensional state through the couplings. We then calculate the evolution of the distances $d_{(t)} = d_0e^{\lambda t}$ between the $3N$-dimensional trajectories and infer the Lyapunov exponents $\lambda$ from the transitioning slope of $\log[d_{(t)}/d_0]$. These are shown in Fig.~\ref{fig:states04}(d) and, as expected, there is a threshold, at about $(J_A = 0.2)$, where the system jumps from regular to chaotic (positive Lyapunov exponent). Additionally, from Eqs.~\eqref{eqS4:eigsSPW1}, we can see that $J_A \in [0, 0.2]$, is a region where the static phase wave I state is also stable, meaning that it is a region of bistability between the static phase wave I and the chaotic regime.

State trajectories in the chaotic regime form  strange attractors whose structures differ according to the parameters $(K, J_A, J_R)$. Fig.~\ref{fig:states05} shows a bidimensional projection of the chaotic trajectory followed by a single particle in the $x-\theta$ torus in four scenarios. We can see that, depending on the scale difference between attracting $(J_A)$ and repulsing $(J_R)$ terms, the trajectories patterns grow from a small disc [Fig.~\ref{fig:states05}(a)], to butterfly-like [Fig.~\ref{fig:states05}(b)], to a disordered single-loop [Fig.~\ref{fig:states05}(c)], and end-up with a scribble-like structure that covers the whole torus [Fig.~\ref{fig:states05}(d)]. 

Another interesting  phenomenon happens when the scale difference between $J_A$ and $J_R$ surpass a threshold, leading to a transition of dimensionality. We can see in the insets of Fig.~\ref{fig:states04}(d) that, despite being inside the chaotic regime, when $(J_A>-2)$, the linear correlation between $x_i$ and $y_i$ is held. However, once $J_A$ gets below this threshold, this correlation breaks. This behavior is strikingly unexpected given that expressions for $\dot{x}_i$ and $\dot{y}_i$ are symmetric, and even the variations in $(J_A, J_R)$ are the same. This dimensionality jump partially explains the scribble-like structure in Fig.~\ref{fig:states05}(d), given that more dimensions might be required to show a clearer attractor structure.

\subsection{Low-dimensional chaos}\label{sub:LDChaos}
\begin{figure*}
	\centering
	\includegraphics[width = \textwidth]{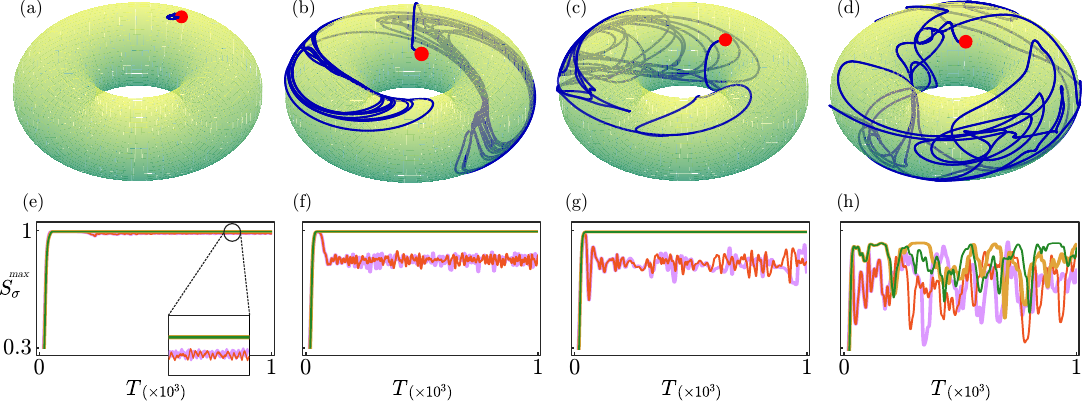}
	\caption{Top row: trajectories (blue) followed by single particles in the $x-\theta$ torus. The red circles are positioned at the start of each trajectory. Bottom row: evolution of $S^{max}_{(\theta\pm x)}$ and $S^{max}_{(x\pm y)}$ for the original initial conditions (purple and gold, respectively) and the perturbed initial conditions (orange and green, respectively). Parameters $(K, J_A, J_R)$ are set as [(a), (e)] $(-1, 0, -0.5)$, [(b), (f)] $(-1,-0.5 ,-0.5)$, [(c), (g)] $(-1, -1,-0.5)$, and [(d), (h)] $(-1, -3,-0.5)$. The small rectangle in (a) shows a magnification of the circled lines. See movies S13, S14, S15, and S16 in the Supplemental Material (for trajectories of two particles see movies S17, S18, S19, and S20).}
	\label{fig:states05}
\end{figure*}

The chaotic behavior of the system can also be identified through the correlation functions defined in Sec~\ref{sub:DS}. Specifically, we use the metrics 
\begin{align*}
	S_{(\theta\pm x)}^{max} &= \max\left[S_{(\theta+x)}, S_{(\theta-x)} \right],\\
	S_{(x\pm y)}^{max} &= \max\left[S_{(x+y)}, S_{(x-y)} \right],
\end{align*}
so we can get a measure of the linear correlation between the respective variables, at any specific time-step, independent on their sign. As shown in the bottom row of Fig.~\ref{fig:states05}, perturbing the initial conditions, generates order parameters that evolve differently. Some of these [Figs.~\ref{fig:states05}(e), (f), (g)] fluctuate around a fixed value, which indicates that we can use them to discern between states. In Fig.~\ref{fig:states05}(h), however, the fluctuations have a much larger amplitude, hindering any attempt to classify the state based on its value. Additionally, the dimensionality jump is also spotted by $S^{max}_{x\pm y}$, given that it converges to $1$ when $J_A$ is over the threshold [Figs.~\ref{fig:states05}(e),(f), and (g)], indicating that $x_i$ and $y_i$ are fully correlated, but it fluctuates once $J_A$ falls below it [Fig.~\ref{fig:states05}(h)].

\section{Attraction-Repulsion phase diagram}\label{sec:sec06}
\begin{figure*}
	\centering
	\includegraphics[width = \textwidth]{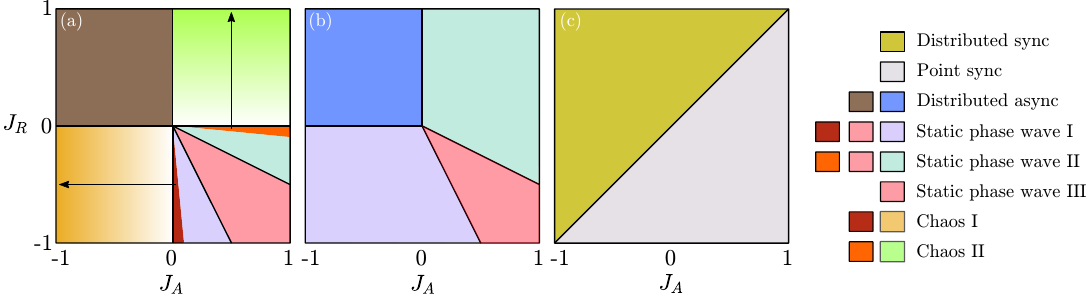}
	\caption{Regions where states emerge on the $J_A - J_R$ plane for (a) $K<0$, (b) $K = 0$, and (c) $K>0$. Arrows and gradients in (a) represent state variations in the chaotic regime, transitioning from swirling to bouncing states. Colors labeling different states are an indicator of multistability.}
	\label{fig:states06}
\end{figure*}

In Fig.~\ref{fig:states06}, we present a summary of the regions where each state emerges in the $J_A - J_R$ plane. These are based on the eigenvalues obtained in Sec.~\ref{sec:sec04} and the Lyapunov exponents calculated in Sec.~\ref{sec:sec05}. 

Note that the active states analyzed in previous sections assumed a positive correlation between $x_i$, $y_i$, and $\theta_i$, corresponding to an extension of static phase wave I for $(K<0)$. We refer to this chaotic regime as Chaos I. Similarly, active effects remain consistent when considering the correlation of static phase wave II, leading us to name that region Chaos II. We can also spot several regions of bistability and multistability. Not only between static states, but also between static and active ones. Remarkably, the regions shown in Fig.~\ref{fig:states06}, consider that initial conditions are distributed randomly in $x$, $y$, and $\theta$. Starting in full synchronization $(\theta_i = \theta)$ and considering frozen phases $(K = 0)$, would lead to the emergence of a synchronous state depending on $(J_A, J_R)$. More exotic behaviors, that we do not study here, can be obtained when considering non-random initial conditions.

\section{Discussion}\label{sec:sec07}
We have studied a swarmalators system where particles move in the 2D plane with periodic boundary conditions. The model, that generalizes the 1D instance introduced in~\cite{o2022collective}, considers attraction and repulsion terms, which ultimately lead the emergence of distributed and active states, some of them similar to the ones introduced in~\cite{o2017oscillators} and~\cite{o2022collective}. All interaction functions involve only sines and cosines, making stability analysis relatively simple, but leading to a weak form of repulsion. In contrast with previous 2D models, we obtained analytical conditions for the emergence of all the static states we presented and also analyzed the nature of the active ones. 

Two states presented in this work, the distributed synchronous and asynchronous states, closely resemble their counterparts described in~\cite{o2017oscillators}, the only difference being that, in our findings, particles are not radially distributed. For these, we determined analytically the conditions that drive their emergence considering only control parameters $(K, J_A, J_R)$. Along this line, we have also introduced an active asynchronous state, that emerges under specific conditions for a finite number of particles. Furthermore, we derived the conditions that guarantee the emergence of a point synchronous state, similar to the one introduced in~\cite{o2022collective} for the 1D model. 

We also described static phase wave states that share features with their 1D~\cite{o2022collective,o2022swarmalators} and 2D counterparts~\cite{o2017oscillators}. The main difference with the former is that, in our model, linear correlations show up in three different ways, as classified in Tab.~\ref{tab:tab01}. Moreover, as pointed before for the distributed synchronous and asynchronous states, our model leads to square, instead of radial, symmetry in particle distribution. Despite these differences, the phase wave states share the \textit{`like attracts like'} feature, characteristic of their definition. Our analytical findings allowed us to state all the conditions in $(K, J_A, J_R)$ that drive the emergence of each type of correlation. Strikingly, we found that for $(K\leq 0)$ there is a region of multistability in the $J_A-J_R$ plane, where the three types of phase wave state could emerge.

Despite the static states and their counterparts found in previous works, the primary distinction between our model and these lies in the active states, which have remained elusive in 1D simplifications of the original swarmalators model. Previous studies where active states emerged in 1D were based on the inclusion of additional parameters, such as frustration~\cite{lizarraga2020synchronization}, external forcing and pinning~\cite{sar2023pinning}, or mixed coupling signs~\cite{hao2023mixed}. Here we demonstrated that active states can emerge solely by adjusting the model parameters $(K, J_A, J_R)$ when repulsion is included. This is coherent with the original idea of swarmalator systems where unsteady states emerge as a result of scale differences between attractive and repulsive effects. Except for the radial symmetry, the active states we found are similar to the splintered and active phase wave states from~\cite{o2017oscillators}.  However, we showed that the active states in our model are part of a spectrum of chaotic states. Even more interesting, we showed that under certain conditions in the chaotic regime, the system suffers a dimensionality transition, where the spatial correlation breaks and particles behave chaotically in $x$, $y$, and $\theta$ axes.

From a theoretical perspective, our model incorporates intriguing characteristics that may pique curiosity as a dynamical system in its own right. Given the system's dimensionality and the somewhat simplistic method we employed to calculate the maximum Lyapunov exponent, it would be worthwhile to explore the entire spectrum of Lyapunov exponents. We suspect that the chaotic regime might exhibit hyperchaotic behaviors under certain conditions, where more than one Lyapunov exponent is positive. It may be also of interest understanding the bifurcations triggered by varying $K$. In Fig~\ref{fig:states06}, we have shown three specific cases that allowed us to discern between the states when $(K < 0)$, $(K = 0)$, and $(K > 0)$. However, it may be worthwhile to study in detail the intermediate states of the system as $K$ transitions from being positive-valued to negative-valued and vice versa. This investigation should provide valuable insights into two open questions: why certain states, like the swirling one, have a very long transient before exhibiting their active characteristics, and what mechanisms underlie the emergence of chaos. Regarding the latter question, we also recommend a rigorous study of the routes to chaos and dimensionality breaking when $(K<0)$. Preliminary numerical studies on this subject have given us hints of emergent intermittency and crises when examining the evolution of the order parameters for different $(J_A, J_R)$.  Finally, it would also be worthwhile to explore whether a threshold on the number of particles $N$ exists, and if so, at what point chaotic behaviors are triggered.

Our model merges two lines of research on the same topic: 1D systems based on the ring model and 2D modifications of the original model. Therefore, its adaptation to previous work on these two lines can occur smoothly. For instance, future work could take our model as a base to consider chirality~\cite{ceron2023diverse}, different phase interactions~\cite{sar2022swarmalators, hong2018active, mclennan2020emergent}, finite cutoff ranges~\cite{lee2021collective}, clusters with different attractive and repulsive strengths~\cite{ghosh2023antiphase}, or short range repulsive interactions~\cite{jimenez2020oscillatory}. Then, given the structure of our model, finding analytical solutions on each of these modifications would be less challenging than doing it for instances of the 2D model introduced in~\cite{o2017oscillators}. 

Given that the interactions in our model are based on Kuramoto-like terms, modifications performed in the Kuramoto model could also be adapted to ours~\cite{acebron2005kuramoto}. A particularly interesting take would be to consider frustration, given that the addition of lag parameters are responsible for the emergence of a turbulent-like state in the ring model~\cite{lizarraga2023synchronization}, and also drive the formation of interesting spatial patterns in swarming only systems~\cite{kruk2018self, kruk2020traveling}. Furthermore, the inclusion of external stimulus discussed in previous work, implies affecting the dynamics in a periodic nature~\cite{sar2023pinning} and mainly in the phase dynamics~\cite{lizarraga2020synchronization, childs2008stability}. One suggestion would be to consider external stimuli in space, mimicking a shear, as has been done for bacterial suspensions~\cite{lopez2015turning}. It would also be of interest considering finite cutoff interaction distances between individuals~\cite{lee2021collective} so that different concentrations of individuals can be managed, emulating the study on vinegar eels in~\cite{quillen2021metachronal}. If a collective of swarmalators works as a medium, it would also be worth studying how it is affected by external particles and vice versa, similar to the study in~\cite{quillen2022fluid} for nematodes. Additionally, we could explore changing the mobility space of the swarmalators, for instance, by imposing discontinuities in the $(x, y)$ space, or also considering self-propulsion of the particles in the direction of their phases, as presented in~\cite{levis2019activity}.

States emerging in our model could also fit collective swarming-only behaviors if we consider that their three spatial degrees of freedom interplay. A clear example of this approach is depicted by the behavior of swarming mosquitoes~\cite{cavagna2023characterization, facchinelli2015stimulating}, which shows similarities with the active asynchronous state in a spherical description. For this setup, we consider switching the model variables such that $\theta_i$ determines the radial distances, and $(x_i, y_i)$, the azimuthal and polar angles, respectively. The resulting state shows clustered individuals in the middle of a cloud of loosely behaved ones, coherent with a description given in~\cite{cavagna2023characterization} (see movie S21 in the Supplemental Material). We also expect the chaotic states, shown in Fig.~\ref{fig:states05}, to be useful in the study of populations that exhibit this type of behaviors (see movie S22 in the Supplemental Material). Finally, our model could be considered as a basis for engineered systems, similar to the ones presented in~\cite{togashi2019modeling, zhou2020coordinating,chen2023nanorobot}.

\begin{acknowledgments} 
	This work was partly supported by FAPESP Grant No. 2021/14335-0 and CNPq Grant No. 301082/2019-7 (M.A.M.A.) and FAPESP Grant No. 2021/04251-4 (J.U.F.L.).
\end{acknowledgments}

\appendix

\section{Derivation of the ring model}
\label{app:app01}
We can collapse the 2D system to a single dimension in space by considering $\dot{x}_i = \dot{y}_i$. Hence, we get rid of the second expression in Eqs.~\eqref{eqS2:expanded}. Additionally, if we assume that there is no scale difference between attraction and repulsion, and that individuals are purely attracting (i.e. $J = J_A = -J_R$, for $J_A>0$), the equations governing the system's dynamics become
\begin{align*}
	\begin{split}
		\dot{x}_i &= \frac{J}{N}\sum_{j = 1}^N\sin(x_{ji})\cos(\theta_{ji}),\\
		\dot{\theta}_i &= \frac{K}{N}\sum_{j = 1}^N \sin(\theta_{ji}) \left[\cos(x_{ji})\right]^2.
	\end{split}
\end{align*}
Notice that, without loss of generality, we have neglected the scaling factor in the phase dynamics ($2K\rightarrow K$). This 1D system fits with the structure described in Eqs.~\eqref{eqS1:1d_general}, and, the only difference with the 1D model introduced in~\cite{o2022collective} is the squared cosine in the phase dynamics. This variation ultimately improves the driving towards synchronous or asynchronous states (depending on the value of $K$), as the influence of the individuals' positions will always be null or positive.

\section{Perturbation analysis of coherent states}\label{app:app02}
For each state described below, we will be able to rearrange the respective perturbation dynamics as 
\begin{equation}
	\delta\dot{\vec{r}}_i~=~\mathbf{R}\delta\vec{r}_i,    
	\label{eqAPPB:GenPertDyn}
\end{equation}
where $\delta\vec{r}_i = \left(\delta x_i, \delta y_i, \delta\theta_i\right)$ and
\begin{equation}
	\mathbf{R} = \begin{pmatrix}
		\mathbf{R_1} & \mathbf{R_2} & \mathbf{R_3}\\
		\mathbf{R_4} & \mathbf{R_5}  & \mathbf{R_6}\\
		\mathbf{R_7} & \mathbf{R_8}  & \mathbf{R_9}
	\end{pmatrix}
	\label{eqAPPB:GenMat}
\end{equation}
is a $3N\times3N$ block matrix where each $N\times N$ block is circulant. Thus, the stability of Eq.~\eqref{eqAPPB:GenPertDyn} can be analyzed by finding the eigenvalues of $\mathbf{R}$. 

We follow the usual procedure to solve an eigenvalue problem, which starts by defining $\mathbf{R_*}~=~\mathbf{R}~-~\vec{\lambda}\mathbf{I}_{3N\times 3N}$ and then concludes by finding the eigenvalues $\vec{\lambda}$ from the determinant of $\mathbf{R_*}$. Thus, considering that $\mathbf{R_*}$ is also a block matrix, we can rewrite its determinant as \[\det\left(\mathbf{R}_*\right) = \det\left(\mathbf{M}\right),\] where
\begin{align*}
	\begin{split}
		\mathbf{M} &= \mathbf{R_1^*}(\mathbf{R_5^*}\mathbf{R_9^*} - \mathbf{R_6}\mathbf{R_8})-\mathbf{R_2}\left(\mathbf{R_4}\mathbf{R_9^*}
		-\mathbf{R_6}\mathbf{R_7}\right)\\ 
		&\qquad  + \mathbf{R_3}\left(\mathbf{R_4}\mathbf{R_8} - \mathbf{R_5^*}\mathbf{R_7}\right),
	\end{split}
\end{align*}
since blocks commute, and
\begin{align*}
	\mathbf{R^*_1} &= \mathbf{R_1}-\vec{\lambda}_{1}\mathbf{I}_N,\\
	\mathbf{R^*_5} &= \mathbf{R_5}-\vec{\lambda}_{5}\mathbf{I}_N,\\
	\mathbf{R^*_9} &= \mathbf{R_9}-\vec{\lambda}_{9}\mathbf{I}_N.\\
\end{align*}
Given that blocks composing $\mathbf{R}$ are circulant, the $N~\times~N$ matrix $\mathbf{M}$ will also be circulant. Then, we can use the general solution to find the determinant of circulant matrices as 
\begin{equation}
	\det\left(\mathbf{M}\right) = \prod_{k = 0}^{N-1}\sum_{r = 0}^{N-1}M_{r+1}\zeta^{kr},
	\label{eqAPPB:GenDet}
\end{equation}
from where we are able to obtain the eigenvalues $\vec{\lambda}$. The term $\zeta = \exp\left\{2\pi\iu/N\right\}$ is a primitive $N$-root of unity. Notice that the structure of $\mathbf{M}$ follows that of $\mathbf{R}$ in Eq.~\eqref{eqAPPB:GenMat} (i.e. elements' sub-indexes represent the same position in the matrix).

\subsection{Point Synchronous}\label{app:PSS}
We add small individual perturbation to the equilibrium states as 
\begin{align*}
	x_i &= x+\delta x_i\\
	y_i &= x+\delta y_i\\
	\theta_i &= x+\delta \theta_i.    
\end{align*}
By plugging these into Eq.~\eqref{eqS2:expanded}, we find the perturbation dynamics, governed by
\begin{align*}
	\begin{split}
		\delta\dot{x}_i &= \frac{J_A - J_R}{2N}\sum_{j= 1}^N(\delta x_j - \delta x_i),\\
		\delta\dot{y}_i &= \frac{J_A - J_R}{2N}\sum_{j= 1}^N(\delta y_j - \delta y_i),\\
		\delta\dot{\theta}_i &= \frac{2K}{N}\sum_{j= 1}^N(\delta \theta_j - \delta\theta_i),
	\end{split}
\end{align*}
which can be arranged as the linear system shown in Eq.~\eqref{eqAPPB:GenPertDyn}. In this particular case, we have that $\left\{\mathbf{R_2}, \mathbf{R_3},\mathbf{R_4},\mathbf{R_6}, \mathbf{R_7} , \mathbf{R_8}\right\} = \mathbf{0}$ and $\mathbf{R_1} = \mathbf{R_5}$, which simplifies the eigenvalue problem considerably. Finally, using Eq.~\eqref{eqAPPB:GenDet}, we obtain the eigenvalues shown in Eqs.~\eqref{eqS4:eigsPSS}.

\subsection{Static phase wave I}\label{app:SPW1}
In this case, the perturbed equilibrium states are defined as
\begin{align*}
	x_i &= \frac{2\pi\iu}{N} + x_0 +\delta x_i,\\
	y_i &= \frac{2\pi\iu}{N} + y_0 +\delta y_i,\\
	\theta_i &= \frac{2\pi\iu}{N} + \theta_0 +\delta \theta_i,
\end{align*}
and the perturbation dynamics are governed by
\begin{align}
	\begin{split}
		\delta\dot{x}_i &=\frac{1}{2N}\sum_j  \left[\delta x_j F_a^{(I)}(i,j)+ \delta\theta_j F_b^{(I)}(i, j)\right]\\
		&\qquad\qquad+\frac{J_R}{2}(\delta x_i - \delta \theta_i),\\
		\delta\dot{y}_i &= \frac{1}{2N}\sum_j\left[\delta y_j F_a^{(I)}(i,j)+\delta\theta_j F_b^{(I)}(i, j)\right]\\
		&\qquad \qquad+ \frac{J_R}{2}(\delta y_i - \delta \theta_i),\\
		\delta\dot{\theta}_i &= \frac{K}{2N}\sum_j \Big[\delta x_j F_c^{(I)}(i,j)
		+\delta y_j F_c^{(I)}(i, j)\\
		& \qquad \qquad +\delta\theta_j F_d^{(I)}(i, j)\Big],
	\end{split}
	\label{eq:posTh_SPW}
\end{align}
where
\begin{align*}
	F_a^{(I)}(i, j) &= J_A\cos\left(\frac{4\pi}{N}(j- i)\right) - J_R,\\
	F_b^{(I)}(i, j) &= J_R + J_A \cos\left(\frac{4\pi}{N}(j-i)\right)\\
	&\qquad \qquad - (J_A + J_R)\cos\left(\frac{2\pi}{N}(j-i)\right) ,\\
	F_c^{(I)}(i, j) &= \cos\left(\frac{6\pi}{N}(j-i)\right) - \cos\left(\frac{2\pi}{N} (j-i)\right),\\
	F_d^{(I)}(i, j) &= \cos\left(\frac{6\pi}{N}(j-i)\right) + 3\cos\left(\frac{2\pi}{N}(j-i)\right).
\end{align*}

The positive correlation between $x_i$ and $y_i$, in the equilibrium, generates a clear symmetry in position dynamics. Then, as noticed, the coefficients $F_a^{(I)}(i, j)$ and $F_b^{(I)}(i, j)$ repeat in the expressions defining $\delta\dot{x}_i$ and $\delta\dot{y}_i$, and $F_c^{(I)}(i, j)$ shows up twice in the definition of $\delta\dot{\theta}_i$. These features allow for the simplification of the linear system when structuring it as in Eq.~\eqref{eqAPPB:GenPertDyn}. Thus, for this case we have that $\mathbf{R_1} = \mathbf{R_5}$, $\mathbf{R_3} = \mathbf{R_6}$, $\mathbf{R_7} = \mathbf{R_8}$, and $\left\{\mathbf{R_2}, \mathbf{R_4}\right\} = \mathbf{0}$. These considerations allow us to split the eigenvalue problem into two problems of lower dimensions such that 
\begin{equation*}
	\mathbf{M}^{(I)} = \mathbf{R_1^*}\left(\mathbf{R_1^*}\mathbf{R_9^*} - 2\mathbf{R_3}\mathbf{R_7}\right).
\end{equation*}
Then, using Eq.~\eqref{eqAPPB:GenDet}, we have that
\begin{widetext}
	\begin{align*}
		\det(\mathbf{R_1^*}) &= \prod_{k = 0}^{N-1} \Bigg\{-\lambda+ \frac{J_R}{2} + \frac{1}{2N}\sum_{j = 1}^{N}\zeta^{(j-1)k}F_a^{(I)}(1, j)\Bigg\},\\
		\det\left(\mathbf{R_1^*}\mathbf{R_9^*} - 2\mathbf{R_3}\mathbf{R_7}\right)  &=  \prod_{k = 0}^{N-1}\Bigg\{\lambda^2 - \lambda \left[\frac{J_R}{2} + \frac{1}{2N}\sum_{j = 1}^N\zeta^{(j-1)k}\left[F_a^{(I)}(1, j) + KF_d^{(I)}(1, j)\right]\right]\\
		& \qquad + \frac{K}{2N}\sum_{j = 1}^N\zeta^{(j-1)k}\left[J_RF_c^{(I)}(1,j) + \frac{J_R}{2}F_d^{(I)}(1,j) - \frac{J_A + J_R}{2}\cos\left(\frac{2\pi}{N}(j-1)\right)\right]\Bigg\},
	\end{align*}
\end{widetext}
whose solution allows us to find the eigenvalues presented in Eqs.~\eqref{eqS4:eigsSPW1}.

\subsection{Static phase wave II}\label{app:SPW2}
The only difference between this state and the previous one, in the equilibrium, is that the correlation between $x_i$ and $\theta_i$ is now negative. Then, after perturbing the equilibrium states individually, we have that
\begin{align*}
	x_i &= \frac{2\pi\iu}{N} + x_0 +\delta x_i,\\
	y_i &= \frac{2\pi\iu}{N} + y_0 +\delta y_i,\\
	\theta_i &= -\frac{2\pi\iu}{N} + \theta_0 +\delta \theta_i,
\end{align*}
and the perturbation dynamics are governed by
\begin{align*}
	\begin{split}
		\delta\dot{x}_i &= \frac{1}{2N}\sum_j\left[\delta x_j F_a^{(II)}(i,j)+\delta\theta_j F_b^{(II)}(i, j)\right]\\
		& \qquad \qquad -\frac{J_A}{2}(\delta x_i + \delta \theta_i) ,\\
		\delta\dot{y}_i &= \frac{1}{2N}\sum_j\left[\delta y_j F_a^{(II)}(i,j)+ \delta\theta_j F_b^{(II)}(i, j)\right]\\
		& \qquad \qquad -\frac{J_A}{2}(\delta y_i + \delta \theta_i),\\
		\delta\dot{\theta}_i &= \frac{K}{2N}\sum_j \Big[\delta x_j F_c^{(II)}(i,j)
		+\delta y_j F_c^{(II)}(i, j)\\
		&\qquad \qquad + \delta\theta_j F_d^{(II)}(i, j)\Big] ,
	\end{split}
\end{align*}
where
\begin{align*}
	F_a^{(II)}(i, j) &= J_A - J_R\cos\left(\frac{4\pi}{N}(j- i)\right),\\
	F_b^{(II)}(i, j) &= J_A + J_R\cos\left(\frac{4\pi}{N}(j-i)\right)\\
	& \qquad \qquad - (J_A + J_R)\cos\left(\frac{2\pi}{N}(j-i)\right),\\
	F_c^{(II)}(i, j) &= \cos\left(\frac{2\pi}{N}(j- i)\right) - \cos\left(\frac{6\pi}{N}(j-i)\right),\\
	F_d^{(II)}(i, j) &= 3\cos\left(\frac{2\pi}{N}(j- i)\right) + \cos\left(\frac{6\pi}{N}(j-i)\right).
\end{align*}
Notice that these equations also show the recurrence of some coefficients in the definition of $\delta \dot{x}_i$, $\delta \dot{y}_i$, and $\delta \dot{\theta}_i$. This consideration leads to the same simplifications described for the previous state when structuring the system as in Eqs.~\eqref{eqAPPB:GenPertDyn}. So we have that, $\mathbf{R_1} = \mathbf{R_5}$, $\mathbf{R_3} = \mathbf{R_6}$, $\mathbf{R_7} = \mathbf{R_8}$, and $\left\{\mathbf{R_2}, \mathbf{R_4}\right\} = \mathbf{0}$.

\begin{equation*}
	\mathbf{M}^{(II)} = \mathbf{R_1^*}\left(\mathbf{R_1^*}\mathbf{R_9^*} - 2\mathbf{R_3}\mathbf{R_7}\right).
\end{equation*}

\begin{widetext}
	\begin{align*}
		\det(\mathbf{R_1^*}) &= \prod_{k = 0}^{N-1} \Bigg\{ -\lambda- \frac{J_A}{2} + \frac{1}{2N}\sum_{j = 1}^{N}\zeta^{(j-1)k}F_a^{(II)}(1, j)\Bigg\},\\
		\det\left(\mathbf{R^*_{1}R^*_{9}} - 2\mathbf{R_3 R_{7}}\right)  &=  \prod_{k = 0}^{N-1}\Bigg\{\lambda^2 - \lambda \left[-\frac{J_A}{2} + \frac{1}{2N}\sum_{j = 1}^N\zeta^{(j-1)k}\left(KF_d^{(II)}(1, j) +F_a^{(II)}(1, j)\right)\right] \\
		& \qquad + \frac{K}{2N}\sum_{j = 1}^N \zeta^{(j-1)k}\left[J_AF_c^{(II)}(1,j) -\frac{J_A}{2}F_d^{(II)}(1,j)+ \frac{(J_A+J_R)}{2}\cos\left(\frac{2\pi}{N}(j-1)\right)\right]\Bigg\},
	\end{align*} 
\end{widetext}
which lead to the eigenvalues presented in Eqs.~\eqref{eqS4:eigsSPW2}.

\subsection{Static phase wave III}\label{app:SPW3}
For this case, the perturbed equilibrium states are defined as
\begin{align*}
	x_i &= \frac{2\pi\iu}{N} + x_0 +\delta x_i,\\
	y_i &= -\frac{2\pi\iu}{N} + y_0 +\delta y_i,\\
	\theta_i &= \frac{2\pi\iu}{N} + \theta_0 +\delta \theta_i,
\end{align*}
and the perturbation dynamics are governed by
\begin{align*}
	\begin{split}
		\delta\dot{x}_i &= \frac{1}{2N}\sum_j\left[\delta x_j F_a^{(III)}(i,j)+\delta\theta_j F_b^{(III)}(i, j) \right]\\
		&\qquad \qquad +\frac{J_R}{2}(\delta x_i - \delta \theta_i),\\
		\delta\dot{y}_i &= \frac{1}{2N}\sum_j \left[\delta y_j F_c^{(III)}(i,j)+\delta\theta_j F_d^{(III)}(i, j)\right]\\
		&\qquad \qquad  -\frac{J_A}{2}(\delta y_i + \delta \theta_i),\\
		\delta\dot{\theta}_i &= \frac{1}{2N}\sum_j \Big[\delta x_j F_e^{(III)}(i,j)
		-\delta y_j F_e^{(III)}(i, j)\\
		&\qquad \qquad + \delta\theta_jF_f^{(III)}(i, j)\Big],
	\end{split}
\end{align*}
where
\begin{widetext}
	\begin{align*}
		F_a^{(III)}(i, j) &= J_A\cos\left(\frac{4\pi}{N}(j- i)\right) - J_R,\\
		F_b^{(III)}(i, j) &= J_R + J_A\cos\left(\frac{4\pi}{N}(j-i)\right) - (J_A + J_R)\cos\left(\frac{2\pi}{N}(j-i)\right),\\
		F_c^{(III)}(i, j) &= J_A - J_R\cos\left(\frac{4\pi}{N}(j- i)\right),\\
		F_d^{(III)}(i, j) &= J_A + J_R\cos\left(\frac{4\pi}{N}(j-i)\right) - (J_A + J_R)\cos\left(\frac{2\pi}{N}(j-i)\right),\\
		F_e^{(III)}(i, j) &= K\cos\left(\frac{6\pi}{N}(j- i)\right) - K\cos\left(\frac{2\pi}{N}(j-i)\right),\\
		F_f^{(III)}(i, j) &= 3K\cos\left(\frac{2\pi}{N}(j- i)\right) + K\cos\left(\frac{6\pi}{N}(j-i)\right).
	\end{align*}
\end{widetext}
Despite the slight difference in the definition of equilibrium states compared to the previous cases (types I and II), simplifications are not possible in the perturbation equations of motion. As noticed, the only repeating coefficient is $F_e^{(III)}(i,j)$ in the definition of $\delta\dot{\theta}_i$. Hence, when structuring the perturbation dynamics as in Eqs.~\eqref{eqAPPB:GenPertDyn}, the only consideration we can make is that $\mathbf{R_7} = - \mathbf{R_8}$. This drawback forces us to write the determinant, using Eq.~\eqref{eqAPPB:GenDet}, as  
\begin{widetext}
	\begin{align*}
		\det(\mathbf{M}^{(III)})&= \prod_{k = 0}^{N-1}\Bigg\{-\lambda^3 + \lambda^2\left[\frac{J_R - J_A}{2} + \frac{1}{2N}S_1\right]+ \lambda\left[\frac{J_AJ_R}{4} +\frac{J_AJ_R}{4N}T_1+ \frac{J_A}{4N}S_2 - \frac{J_R}{4N}S_3 \right] \\
		&\qquad \qquad+ \frac{K}{16N}\left(J_A + J_R\right)^2 - \frac{J_AJ_R}{8N}S_4\Bigg\},
	\end{align*}
\end{widetext}
where
\small
\begin{align*}
	S_1 &= \sum_{j = 1}^{N}\zeta^{(j-1)k}\Big(F^{(III)}_a(1,j) + F^{(III)}_c(1,j) + KF^{(III)}_f(1,j)\Big),
\end{align*}
\begin{align*}
	S_2 &= \sum_{j = 1}^N\zeta^{(j-1)k}\Big(F_a^{(III)}(1,j) + KF_e^{(III)}(1,j) + KF_f^{(III)}(1,j)\Big),
\end{align*}    
\begin{align*}
	S_3 &= \sum_{j = 1}^N\zeta^{(j-1)k}\Big(F_c^{(III)}(1,j) + KF_e^{(III)}(1,j) + KF_f^{(III)}(1,j)\Big),
\end{align*}
\normalsize
\begin{align*}
	S_4 &= \sum_{j = 1}^N\zeta^{(j-1)k}\left(2F_e^{(III)}(1,j) + F_f^{(III)}(1,j)\right),\\
	T_1 &= \sum_{j = 1}^N\zeta^{(j-1)k} + 0.25\left(\zeta^{(j-1)(k+2)} + \zeta^{(j-1)(k-2)}\right),\\
	T_2 &= 0.5\sum_{j = 1}^k \zeta^{(j-1)(k+1)} + \zeta^{(j-1)(k-1)}.
\end{align*}
To find the eigenvalues, we must address $N$ cubic equations of the form $-\lambda^3 + \lambda^2 \beta+ \lambda \xi  + \eta$. We solve these using the general formula for 
\begin{equation}
	\beta = 
	\begin{cases}
		0 ,& \text{for } k = 0;\\   
		0.5(J_R - J_A) + 1.5K ,& \text{for } k = 1, N-1;\\   
		0.25(J_R - J_A) ,& \text{for } k = 2, N-2;\\   
		0.5(J_R - J_A) + 0.25K ,& \text{for } k = 3, N-3;\\
		0.5(J_R - J_A) ,& \text{otherwise};\\   
	\end{cases}
\end{equation}
\begin{equation}
	\xi = 
	\begin{cases}
		0 ,& \text{for } k = 0,\\   
		0.25(J_AJ_R + K(J_A - J_R)) ,& \text{for } k = 1, N-1;\\   
		0.125(J_A^2 + J_R^2 + 2.5J_AJ_R) ,& \text{for } k = 2, N-2;\\   
		0.25(J_AJ_R + K(J_A - J_R)) ,& \text{for } k = 3, N-3;\\
		0.25KJ_AJ_R ,& \text{otherwise};\\   
	\end{cases}
\end{equation}
and
\begin{equation*}
	\eta = 
	\begin{cases}
		0 ,& \text{for } k = 0;\\   
		0.125K[0.25(J_A+J_R)^2 & {}\\
		\qquad -J_AJ_R] ,& \text{for } k = 1, N-1;\\
		0 ,& \text{for } k = 2, N-2;\\   
		-0.1875KJ_AJ_R,& \text{for } k = 3, N-3;\\
		0,& \text{otherwise}.\\   
	\end{cases}
\end{equation*}

\section{Perturbation analysis of incoherent states}\label{app:app03}
In general, we perturb the equilibrium state $\rho_0$ by a small quantity $\delta\rho$, such that $\rho = \rho_0 + \delta\rho$. Then, the temporal evolution of the perturbation is governed by
\begin{equation}
	\frac{\partial}{\partial t}\delta\rho = -\vec{\nabla}(\delta\rho)\vec{v}.
	\label{eq:pert_inc_dyn}
\end{equation}
From Eq.~\eqref{eqS4:norm}, we know that 
\begin{equation*}
	\int_{0}^{2\pi}\delta\rho(x, y, \theta, t)\mathrm{d}x\mathrm{d}y\mathrm{d}\theta = 0,
\end{equation*}
and to first order in $\delta\rho(x, \theta, t)$, we have that
\begin{equation}
	S_{\sigma}^1e^{\iu\phi_{\sigma}} = \int_{0}^{2\pi}e^{\iu\sigma}\delta\rho(x, y, \theta, t)\mathrm{d}x\mathrm{d}y\mathrm{d}\theta.
	\label{eq:firstOP}
\end{equation}

We expand $\delta\rho$ in Fourier series as
\begin{equation}
	\delta\rho = \sum_{m,n,l}f_{m,n,l}(t)e^{\iu(mx+ny+l\theta)},
	\label{eq:fourier}
\end{equation}
and we will solve equations of the type $f(t) = \bar{f}e^{\lambda t}$ to get the eigenvalues $\lambda$.

\subsection{Distributed asynchronous}\label{app:DA}
We consider the equilibrium $\rho_0^{A}$ defined in Eqs.~\eqref{eqS4:incohEq}. Using Eqs.~\eqref{eq:pert_inc_dyn} and~\eqref{eq:firstOP}, we obtain the perturbation dynamics 

\begin{widetext}
	\begin{align}
		\begin{split}
			\frac{d}{dt}\delta\rho &=\frac{J_A}{16\pi^3}\left[S_{(\theta + x)}^1\cos\left(\phi_{(\theta + x)} - \theta - x\right) + S_{(\theta + y)}^1\cos\left(\phi_{(\theta + y)} - \theta - y\right)\right]\\
			&\qquad- \frac{J_R}{16\pi^3}\left[S_{(\theta - x)}^1\cos\left(\phi_{(\theta - x)} - \theta + x\right) + S_{(\theta - y)}^1\cos\left(\phi_{(\theta - y)} - \theta + y\right)\right]\\
			& \qquad \qquad + \frac{K}{32\pi^3}\bigg[S_{(\theta + 2x)}^1\cos\left(\phi_{(\theta + 2x)} - \theta - 2x\right) + S_{(\theta - 2x)}^1\cos\left(\phi_{(\theta - 2x)} - \theta + 2x\right)\\
			& \qquad \qquad \qquad + S_{(\theta + 2y)}^1\cos\left(\phi_{(\theta + 2y)} - \theta - 2y\right)+ S_{(\theta - 2y)}^1\cos\left(\phi_{(\theta - 2y)} - \theta + 2y\right)+ 4S_{\theta}^1\cos\left(\phi_{\theta} - \theta\right)\bigg]. 
		\end{split}
	\end{align}
\end{widetext}
We expand this equation following the form of Eq.~\eqref{eq:fourier} and we will see that the only relevant terms correspond to 

\begin{align}
	\begin{split}
		\{(m, n, l)\} &= \{(-1, 0, 1); (0, -1, -1); 
		(1, 0, -1);\\ 
		&\qquad \quad(0, 1, -1);
		(0, 0, -1); (2, 0, -1); \\&\qquad \qquad(-2, 0, -1);
		(0, 2, -1); (0, -2, -1)\}.
	\end{split}
\end{align}
Then, the solutions give us the eigenvalues shown in Eqs.~\eqref{eqS4:eigs_SA}.

\subsection{Distributed synchronous}\label{app:DS}
Given that in this state phases are synchronized and only positions are uniformly distributed, we uncouple the system's dynamics. The incoherence between phases and positions will lead to zeroing order parameters $S_{(\theta\pm2x)}$ and $S_{(\theta\pm2y)}$. That allows us to rewrite \[v_{\theta} = KS_{\theta}\sin(\phi_{\theta}- \theta),\] which is exactly the mean-field description of the Kuramoto model without considering natural frequencies~\cite{strogatz2000kuramoto}. Then, the condition to reach synchronization is given by $(K>0)$. 

In the equilibrium, the synchronization of phases will lead to $\theta_i = \theta$. This condition allows for the equivalences
\begin{align*}
	\begin{split}
		S_{(\theta \pm x)}e^{\iu\phi_{(\theta \pm x)}} &=S_xe^{\iu(\theta \pm\phi_x)}, \\
		S_{(\theta \pm y)}e^{\iu\phi_{(\theta \pm y)}} &=S_ye^{\iu(\theta \pm\phi_y)},
	\end{split}
\end{align*}
and therefore, the velocity field for this state is defined as  
\begin{align}
	\begin{split}
		v_x &= \frac{J_A- J_R}{2}S_x\sin\left(\phi_x - x\right),\\
		v_y &= \frac{J_A- J_R}{2}S_y\sin\left(\phi_y - y\right).
	\end{split}
\end{align}
In this case we use the equilibrium $\rho_0^S$ from Eqs.~\eqref{eqS4:incohEq}, and considering ~\eqref{eq:firstOP}, the perturbation dynamics are governed by 
\begin{equation}
	\frac{d}{dt}\delta\rho =\frac{J_A - J_R}{8\pi^2}\left[S_x^1\cos\left(\phi_x - x\right)+ S_y^1\cos\left(\phi_y - y\right)\right]. 
\end{equation}
Then, using Eq.~\eqref{eq:fourier} for this equation, we find the eigenvalue shown in Eq.~\eqref{eqS4:eigs_SS}.
\nocite{*}


\providecommand{\noopsort}[1]{}\providecommand{\singleletter}[1]{#1}%

\end{document}